
\font\fiverm=cmr5
\font\fivebf=cmbx5
\font\fivei=cmmi5

\font\fivesy=cmsy5

\font\sevenrm=cmr7
\font\sevenbf=cmbx7
\font\seveni=cmmi7

\font\sevensy=cmsy7

\font\ninerm=cmr9
\font\ninebf=cmbx9

\font\nineit=cmmi9
\font\ninei=cmmi9 
\font\ninesy=cmsy9  
\font\nineex=cmex10
\font\tenrm=cmr10
\font\tenbf=cmbx10
\font\tensl=cmsl10
\font\tenit=cmmi10
\font\teni=cmmi10 
 
\font\tensy=cmsy10
\font\tenex=cmex10
\font\twelverm=cmr12
\font\twelvebf=cmbx12
\font\twelvesl=cmsl12
\font\twelveit=cmmi12
\font\twelvei=cmmi12
\font\twelvesy=cmsy10 scaled\magstep1


%
%
 %
%
 \def\ninepoint{%
   \normalbaselineskip=11pt
   \def\rm{\fam0\ninerm}%
   \def\it{\fam0\nineit}%
   \def\bf{\fam\bffam\ninebf}%
   \def\bi{\fam\bffam\ninebf}%
   \def\rmit{\fam0\ninerm\def\it{\fam0\nineit}}%
   \def\bfit{\fam\bffam\ninebf\def\it{\bi}}%
   \def\bsl{\fam\bffam\ninebsl}
   \textfont0=\ninerm\scriptfont0=\sevenrm\scriptscriptfont0=\fiverm
   \textfont1=\ninei\scriptfont1=\seveni\scriptscriptfont1=\fivei
   \textfont2=\ninesy\scriptfont2=\sevensy\scriptscriptfont2=\fivesy
   \textfont3=\nineex \scriptfont3=\tenex \scriptscriptfont3=\tenex
   \textfont\bffam=\ninebf\scriptfont\bffam=\sevenbf\scriptscriptfont\bffam=
     \fivebf
   \normalbaselines\rm}%
%
 \def\tenpoint{%
   \normalbaselineskip=12pt
   \def\rm{\fam0\tenrm}%
   \def\it{\fam0\tenit}%
   \def\bf{\fam\bffam\tenbf}%
   \def\bi{\fam\bffam\tenbf}%
   \def\rmit{\fam0\tenrm\def\it{\fam0\tenit}}%
   \def\bfit{\fam\bffam\tenbf\def\it{\bi}}%
   \def\bsl{\fam\bffam\tenbsl}
   \textfont0=\tenrm\scriptfont0=\sevenrm\scriptscriptfont0=\fiverm
   \textfont1=\teni\scriptfont1=\seveni\scriptscriptfont1=\fivei
   \textfont2=\tensy\scriptfont2=\sevensy\scriptscriptfont2=\fivesy
   \textfont3=\tenex \scriptfont3=\tenex \scriptscriptfont3=\tenex
     \fivebf
   \textfont\bffam=\tenib\scriptfont\bffam=\sevenib\scriptscriptfont\bffam=
     \fiveib
   \normalbaselines\rm}%
%

%

     
\font\twelverm=cmr10 scaled 1200    \font\twelvei=cmmi10 scaled 1200
\font\twelvesy=cmsy10 scaled 1200   \font\twelveex=cmex10 scaled 1200
\font\twelvebf=cmbx10 scaled 1200   \font\twelvesl=cmsl10 scaled 1200
\font\twelvett=cmtt10 scaled 1200   \font\twelveit=cmti10 scaled 1200
     
\skewchar\twelvei='177   \skewchar\twelvesy='60
     
     
\def\twelvepoint{\normalbaselineskip=12.4pt
  \abovedisplayskip 12.4pt plus 3pt minus 9pt
  \belowdisplayskip 12.4pt plus 3pt minus 9pt
  \abovedisplayshortskip 0pt plus 3pt
  \belowdisplayshortskip 7.2pt plus 3pt minus 4pt
  \smallskipamount=3.6pt plus1.2pt minus1.2pt
  \medskipamount=7.2pt plus2.4pt minus2.4pt
  \bigskipamount=14.4pt plus4.8pt minus4.8pt
  \def\rm{\fam0\twelverm}          \def\it{\fam\itfam\twelveit}%
  \def\sl{\fam\slfam\twelvesl}     \def\bf{\fam\bffam\twelvebf}%
  \def\mit{\fam 1}                 \def\cal{\fam 2}%
  \def\tt{\twelvett}
  \textfont0=\twelverm   \scriptfont0=\tenrm   \scriptscriptfont0=\sevenrm
  \textfont1=\twelvei    \scriptfont1=\teni    \scriptscriptfont1=\seveni
  \textfont2=\twelvesy   \scriptfont2=\tensy   \scriptscriptfont2=\sevensy
  \textfont3=\twelveex   \scriptfont3=\twelveex  \scriptscriptfont3=\twelveex
  \textfont\itfam=\twelveit
  \textfont\slfam=\twelvesl
  \textfont\bffam=\twelvebf \scriptfont\bffam=\tenbf
  \scriptscriptfont\bffam=\sevenbf
  \normalbaselines\rm}
     
     
\def\tenpoint{\normalbaselineskip=12pt
  \abovedisplayskip 12pt plus 3pt minus 9pt
  \belowdisplayskip 12pt plus 3pt minus 9pt
  \abovedisplayshortskip 0pt plus 3pt
  \belowdisplayshortskip 7pt plus 3pt minus 4pt
  \smallskipamount=3pt plus1pt minus1pt
  \medskipamount=6pt plus2pt minus2pt
  \bigskipamount=12pt plus4pt minus4pt
  \def\rm{\fam0\tenrm}          \def\it{\fam\itfam\tenit}%
  \def\sl{\fam\slfam\tensl}     \def\bf{\fam\bffam\tenbf}%
  \def\smc{\tensmc}             \def\mit{\fam 1}%
  \def\cal{\fam 2}%
  \textfont0=\tenrm   \scriptfont0=\sevenrm   \scriptscriptfont0=\fiverm
  \textfont1=\teni    \scriptfont1=\seveni    \scriptscriptfont1=\fivei
  \textfont2=\tensy   \scriptfont2=\sevensy   \scriptscriptfont2=\fivesy
  \textfont3=\tenex   \scriptfont3=\tenex     \scriptscriptfont3=\tenex
  \textfont\itfam=\tenit
  \textfont\slfam=\tensl
  \textfont\bffam=\tenbf \scriptfont\bffam=\sevenbf
  \scriptscriptfont\bffam=\fivebf
  \normalbaselines\rm}
     

{\obeylines\gdef\
{}}
\def\singlespace{\baselineskip=\normalbaselineskip}

\def\doublespace{\baselineskip=\normalbaselineskip \multiply\baselineskip by 2}

\newcount\firstpageno
\firstpageno=2
\footline={\ifnum\pageno<\firstpageno{\hfil}\else{\hfil\twelverm\folio\hfil}\fi}
\let\rawfootnote=\footnote              
\def\footnote#1#2{{\rm\singlespace\parindent=0pt\rawfootnote{#1}{#2}}}

     
\hsize=6.5truein
\hoffset=0truein
\vsize=8.9truein
\voffset=0truein
\parskip=\medskipamount
\twelvepoint            
\doublespace            
\overfullrule=0pt       
     
     
\def\preprintno#1{
 \rightline{\rm #1}}    
     
     
\def\ref#1{Ref. #1}                     

\def\frac#1#2{{\textstyle{#1 \over #2}}}
\def\half{{\textstyle{ 1\over 2}}}

\def\sla{\raise.15ex\hbox{$/$}\kern-.57em}
\def\leaderfill{\leaders\hbox to 1em{\hss.\hss}\hfill}
\def\twiddle{\lower.9ex\rlap{$\kern-.1em\scriptstyle\sim$}}
\def\bigtwiddle{\lower1.ex\rlap{$\sim$}}
\def\gtwid{\mathrel{\raise.3ex\hbox{$>$\kern-.75em\lower1ex\hbox{$\sim$}}}}
\def\ltwid{\mathrel{\raise.3ex\hbox{$<$\kern-.75em\lower1ex\hbox{$\sim$}}}}
\def\square{\kern1pt\vbox{\hrule height 1.2pt\hbox{\vrule width 1.2pt\hskip 3pt
   \vbox{\vskip 6pt}\hskip 3pt\vrule width 0.6pt}\hrule height 0.6pt}\kern1pt}

\def\Fint{\rlap{$\Biggl\rfloor$}\Biggl\lceil}

\def\m@th{\mathsurround=0pt }
\def\leftrightarrowfill{$\m@th \mathord\leftarrow \mkern-6mu
 \cleaders\hbox{$\mkern-2mu \mathord- \mkern-2mu$}\hfill
 \mkern-6mu \mathord\rightarrow$}
\def\overleftrightarrow#1{\vbox{\ialign{##\crcr
     \leftrightarrowfill\crcr\noalign{\kern-1pt\nointerlineskip}
     $\hfil\displaystyle{#1}\hfil$\crcr}}}

\input psfig.sty
\singlespace
\preprintno{hep-ph/9602315}
\preprintno{CPTH-S372.0995}
\preprintno{CRETE-95-11}
\preprintno{UFIFT-HEP-95-17}
\preprintno{Revised June, 1996}
\vskip 2cm
\centerline{\bf QUANTUM GRAVITY SLOWS INFLATION}
\vskip 2cm
\centerline{\bf N. C. Tsamis$^{*}$}
\vskip .5cm
\centerline{\it Centre de Physique Th\'eorique, Ecole Polytechnique}
\centerline{\it Palaiseau 91128, FRANCE}
\vskip .5cm
\centerline{and}
\vskip .5cm
\centerline{\it Theory Group, FO.R.T.H.}
\centerline{\it Heraklion, Crete 71110, GREECE}
\vskip 1cm
\centerline{and}
\vskip 1cm
\centerline{\bf R. P. Woodard$^{\dagger}$}
\vskip .5cm
\centerline{\it Department of Physics, University of Florida}
\centerline{\it Gainesville, FL 32611, USA}
\vskip 2cm
\centerline{ABSTRACT}
\itemitem{}{\tenpoint We consider the quantum gravitational back-reaction on 
an initially inflating, homogeneous and isotropic universe whose topology is 
$T^3 \times \Re$. Although there is no secular effect at one loop, an explicit
calculation shows that two-loop processes act to slow the rate of expansion by
an amount which becomes non-perturbatively large at late times. By exploiting
Feynman's tree theorem we show that all higher loops act in the same sense.}
\footnote{}{$^*$~~ \tenpoint e-mail: tsamis@iesl.forth.gr and 
tsamis@orphee.polytechnique.fr}
\footnote{}{$^{\dagger}$~~ \tenpoint e-mail: woodard@phys.ufl.edu}

\vfill\eject
\doublespace

\centerline {\bf 1. Introduction}

Inflation provides a wonderful explanation for the fact that the cosmic 
microwave background is observed to be in thermal equilibrium to about one 
part in $10^5$, even for regions of the universe which are only now coming 
into contact with one another [1]. However, inflationary cosmology has no 
pretensions of explaining why the dimensionless product of the cosmological 
constant and Newton's constant is observed to be zero to within about one 
part in $10^{120}$ [2]. Indeed, the assumption of inflation imposes severe 
restrictions on any proposal for understanding this grotesque hierarchy. 
For if we assume that the responsible mechanism can screen a cosmological 
constant of any size, and that it makes no distinction between the bare 
cosmological constant and contributions from the matter potential, then 
one has to explain why inflation was ever able to start. One must also 
understand why the screening mechanism operates slowly enough to permit 
the $\approx 55$ e-foldings of inflation needed to produce the isotropy 
of the cosmic microwave background.

We have proposed that the cosmological constant isn't 
unreasonably small but only appears so due to the screening effect of 
infrared processes in quantum gravity [3,4].
\footnote{*}{\tenpoint It has also been suggested that screening can 
occur due to Hawking radiation [5], or from the existence of an infrared 
fixed point in various effective theories of gravitation [6]. Our 
mechanism is nearest to Ford's proposal [7], which was based on the 
assumption that the coincidence limit of the graviton propagator grows 
in time. Note, however, that our formalism does not show temporal 
growth for the coincident propagator. It is also significant that --- 
if such growth had been present --- Ford's effect would emerge from 
an entirely different set of diagrams than ours.} 
These infrared processes can become strong because the graviton is 
massless and because a non-zero cosmological constant endows it with 
a self-interaction of dimension three. (By way of contrast, the massless 
gluons of QCD possess only dimension four couplings.) Since dimension 
three couplings between positive norm bosons lower the vacuum energy, 
the effect is to screen the bare cosmological constant. Like all infrared 
effects, our process derives from the causal and coherent superposition 
of interactions throughout the past lightcone. The effect is absent 
before the onset of inflation because thermal fluctuations disrupt the 
coherent superposition of interactions from different regions of the 
very early universe; it is only after inflation has redshifted the 
temperature that a coherent effect can begin to accumulate. This effect 
eventually becomes arbitrarily strong because the invariant volume of 
the past lightcone from the onset of inflation grows without bound as the
future unfolds. There is a long period of inflation because an enormous
invariant volume is needed to overcome the natural weakness of gravitational 
interactions.

Since our effect comes from the infrared we can quantize Einstein's theory:
$${\cal L} = {1 \over 16 \pi G} \Bigl(R - 2 \Lambda\Bigr) \; 
\sqrt{-g} + {\rm (counterterms)} \eqno(1)$$
without worrying about the still unknown corrections which must be added 
to avoid inconsistencies on the Planck scale. The modes which contribute 
most strongly at any time turn out to have physical wavelengths of about 
the Hubble radius. As long as the scale of inflation is a few orders of 
magnitude below the Planck mass, modes which redshift down from the unknown
ultraviolet sector will have plenty of time to reach an equilibrium governed
by (1).

Another significant feature of our mechanism is its uniqueness to gravity. 
Interactions mediated by massive quanta cannot give a strong infrared effect 
because they do not superpose coherently. Conformally invariant quanta cannot
give a strong effect because they are insensitive to the enormous invariant 
volume in the conformally flat geometry created by a long period of 
homogeneous and isotropic inflation. The graviton is unique among known 
particles in being massless but not conformally invariant.

A necessary consequence of our mechanism is that asymptotic quantum field 
theory must break down if one makes the incorrect assumption that the ``out'' 
vacuum shows inflation. Previous explicit calculations have confirmed this, 
both for ``in''-``out'' matrix elements [3,4], and for scattering amplitudes
[8]. These results imply that corrections to the background must become 
non-perturbatively large at late times, but they do not fix the rate at 
which this occurs. To determine this rate it is necessary to follow the 
evolution of the expectation value of the background in the presence of 
a plausible initial state. Our proposal for such a calculation has been 
discussed elsewhere at great length [4] and we shall content ourselves here 
with a brief review of the formalism in section 2. The main point of this 
paper is to announce that we have brought the calculation to a successful 
conclusion (section 3) and to discuss the result (section 4).

\vskip 1cm
\centerline {\bf 2. The formalism}

Because it is unlikely for inflation to begin simultaneously over more than 
a small region, we work on the manifold $T^3 \times \Re$, with the physical 
distances of the toroidal radii equal to a Hubble length at the onset of 
inflation. The object of our study is the expectation value of the invariant 
element in the presence of a state which is initially free de Sitter vacuum. 
Since the state is homogeneous and isotropic the result can be expressed in 
co-moving coordinates:
$$\Bigl\langle 0 \Bigl\vert \; g_{\mu \nu}(t,{\vec x}) \;
dx^{\mu} dx^{\nu} \; \Bigr\vert 0 \Bigr\rangle = 
-dt^2 + {\rm a}^2(t) d{\vec x} \cdot d{\vec x} \eqno(2)$$
We take the onset of inflation to be $t=0$, and we work perturbatively 
around the classical background:
$${\rm a}_{\rm class}(t) = \exp(Ht) \eqno(3)$$
where the Hubble constant is $H \equiv \sqrt{\frac13 \Lambda}$. The actual 
rate of expansion is given by the effective Hubble constant:
$$H_{\rm eff}(t) \equiv {d \ln({\rm a}) \over dt} \eqno(4)$$
which is an invariant by virtue of having been defined in a unique coordinate
system.

It is simplest to perform the calculation in conformally flat coordinates, 
for which the invariant element of the background is:
$$-dt^2 + {\rm a}^2_{\rm class}(t) \; d{\vec x} \cdot d{\vec x} = \Omega^2
\Bigl(-du^2 + d{\vec x} \cdot d{\vec x}\Bigr) \eqno(5a)$$
$$\Omega \equiv {1 \over H u} = \exp(H t) \eqno(5b)$$
Note the temporal inversion and the fact that the onset of inflation at 
$t=0$ corresponds to $u = H^{-1}$. Since the infinite future is at $u = 
0^+$, and since the spatial coordinates fall within the region, $-\frac12 
H^{-1} < x^i \leq \frac12 H^{-1}$, the range of conformal coordinates is 
rather small. This is why a conformally invariant field --- whose dynamics 
are locally the same as in flat space, except for ultraviolet regularization 
--- cannot induce a big infrared effect.

Perturbation theory is organized most conveniently in terms of a 
``pseudo-graviton'' field, $\psi_{\mu \nu}$, obtained by conformally 
re-scaling the metric:
$$g_{\mu \nu} \equiv \Omega^2 \; {\widetilde g}_{\mu \nu} \equiv 
\Omega^2 \; \Bigl(\eta_{\mu \nu} + \kappa \psi_{\mu \nu}\Bigr) 
\eqno(6)$$
As usual, pseudo-graviton indices are raised and lowered with the Lorentz
metric, and the loop counting parameter is $\kappa^2 \equiv 16 \pi G$. 
After some judicious partial integrations the invariant part of the bare 
Lagrangian takes the following form [9]:
$$\eqalignno{{\cal L}_{\rm inv} =
\sqrt{-{\widetilde g}} \; {\widetilde g}^{\alpha \beta} \;
{\widetilde g}^{\rho \sigma} \; {\widetilde g}^{\mu \nu} 
\Bigl[\half \psi_{\alpha \rho , \mu} \; \psi_{\nu \sigma , \beta} - 
\half \psi_{\alpha \beta , \rho} &\; \psi_{\sigma \mu , \nu} + 
\frac14 \psi_{\alpha \beta , \rho} \; \psi_{\mu \nu , \sigma} - 
\frac14 \psi_{\alpha \rho , \mu} \; \psi_{\beta \sigma , \nu}\Bigr] 
\Omega^{2} \cr &- \half \sqrt{-{\widetilde g}} \; 
{\widetilde g}^{\rho \sigma} \; {\widetilde g}^{\mu \nu} \; 
\psi_{\rho \sigma , \mu} \; \psi_{\nu}^{~\alpha} \; 
(\Omega^{2})_{,\alpha} &(7) \cr}$$
Note that each interaction term contains at least one ordinary derivative. 
This occurs because the dimension three coupling is canceled by the 
undifferentiated terms from the covariant derivatives of the dimension 
five coupling. Such a cancellation --- for which there is no scalar field 
or flat space analog --- is essential for classical stability [10] against 
growth of zero modes. An interesting consequence is that the leading 
infrared effects cancel as well in the quantum theory. However, the two 
couplings do not agree at subleading order, and there is still a very 
strong quantum effect.

Gauge fixing is accomplished through the addition of $-\half \eta^{\mu \nu} 
F_{\mu} F_{\nu}$ where [9]:
$$F_{\mu} \equiv \Bigl(\psi^{\rho}_{~\mu , \rho} - 
\frac12 \psi^{\rho}_{~\rho , \mu} + 2 \psi^{\rho}_{~\mu} \; 
{(\ln \Omega)}_{,\rho}\Bigr) \Omega \eqno(8)$$
The associated ghost Lagrangian is [9]:
$$\eqalignno{{\cal L}_{\rm ghost} = -\Omega^2 \; 
&{\overline \omega}^{\mu , \nu} \; \Bigl[{\widetilde g}_{\rho \mu} \; 
\partial_{\nu} + {\widetilde g}_{\rho \nu} \; \partial_{\mu} + 
{\widetilde g}_{\mu \nu , \rho} + 2 {\widetilde g}_{\mu \nu} \; 
{(\ln \Omega)}_{, \rho} \Bigr] \; \omega^{\rho} \cr 
&+ {\Bigl( \Omega^2 \; {\overline \omega}^{\mu} \Bigr)}_{, \mu} 
\eta^{\rho \sigma} \; \Bigl[{\widetilde g}_{\nu \rho} \; 
\partial_{\sigma} + \frac12 {\widetilde g}_{\rho \sigma , \nu} + 
{\widetilde g}_{\rho \sigma} \; {(\ln \Omega)}_{, \nu} \Bigr] \; 
\omega^{\nu} &(9) \cr}$$
The zeroth order action results in the following free field expansion [11]:
$$\psi_{\mu \nu}(u,{\vec x}) = 
\Biggl({{\rm Zero} \atop {\rm Modes}}\Biggr) + 
H^3 \sum_{\lambda, {\vec k}\neq 0} \Biggl\{ \Psi_{\mu \nu}\Bigl(u,{\vec x};
{\vec k},\lambda\Bigr) \; a({\vec k},\lambda) + \Psi_{\mu \nu}^*
\Bigl(u,{\vec x};{\vec k},\lambda\Bigr) \; a^{\dagger}({\vec k},\lambda) 
\Biggr\} \eqno(10)$$
The spatial polarizations consist of ``A'' modes:
$$\Psi_{\mu \nu}\Bigl(u,{\vec x};{\vec k},\lambda\Bigr) = 
{Hu \over \sqrt{2 k}} \; \Bigl(1 + {i \over k u}\Bigr) \; 
\exp\Bigl[i k \Bigl(u - \frac1{H}\Bigr) + 
i {\vec k} \cdot {\vec x}\Bigr] \; \epsilon_{\mu \nu}({\vec k},\lambda) 
\qquad \forall \lambda \in A \eqno(11a)$$
while the space--time and purely temporal polarizations are associated, 
respectively, with ``B'' and ``C'' modes:
$$\Psi_{\mu \nu}\Bigl(u,{\vec x};{\vec k},\lambda\Bigr) = 
{Hu \over \sqrt{2 k}} \; \exp\Bigl[i k \Bigl(u - \frac1{H}\Bigr) + 
i {\vec k} \cdot {\vec x} \Bigr] \; \epsilon_{\mu \nu}({\vec k},\lambda) 
\qquad \forall \lambda \in B,C \eqno(11b)$$
In LSZ reduction one would integrate against and contract into $\Psi_{\mu 
\nu}(u,{\vec x};{\vec k},\lambda)$ to insert and ``in''-coming graviton of 
momentum ${\vec k}$ and polarization $\lambda$; the conjugate would be used 
to extract an ``out''-going graviton with the same quantum numbers. The 
zero modes evolve as free particles with time dependences $1$ and $u^3$ 
for the A modes, and $u$ and $u^2$ for the B and C modes. Since causality 
decouples the zero modes shortly after the onset of inflation, they play 
no role in screening and we shall not trouble with them further.

We define $\vert 0 \rangle$ as the Heisenberg state annihilated by 
$a({\vec k}, \lambda)$ --- and the analogous ghost operators --- at 
the onset of inflation. We can use this condition and expansion (10) 
to express the free pseudo-graviton propagator as a mode sum [8]:
$$\eqalignno{i\Bigl[ {_{\mu \nu}}\Delta_{\rho \sigma}\Bigr](x;x') &\equiv
\Bigl\langle 0 \Bigl\vert T\Bigl\{\psi_{\mu \nu}(x) \; 
\psi_{\rho \sigma}(x')\Bigr\} \Bigr\vert 0 \Bigr\rangle_{\rm free} 
&(12a) \cr 
&= H^3 \sum_{\lambda, {\vec k} \neq 0} \Biggl\{ \theta(u'-u) \; \Psi_{\mu \nu}
\; {\Psi'}^*_{\rho \sigma} + \theta(u-u') \; \Psi^*_{\mu \nu} \; {\Psi'}_{\rho 
\sigma} \Biggr\} e^{- \epsilon \Vert {\vec k} \Vert} &(12b) \cr}$$
Note that the convergence factor $e^{- \epsilon \Vert {\vec k} \Vert}$ serves 
as an ultraviolet mode cutoff. Although the resulting regularization is very 
convenient for this calculation, its failure to respect general coordinate 
invariance necessitates the use of non-invariant counterterms. These are 
analogous to the photon mass which must be added to QED when using a momentum 
cutoff. Just as in QED, these non-invariant counterterms do not affect long 
distance phenomena.

Because the propagator is only needed for small conformal coordinate 
separations, ${\Delta x} \equiv \Vert {\vec x}' - {\vec x} \Vert$ and 
${\Delta u} \equiv u' - u$, the sum over momenta is well approximated as 
an integral. When this is done the pseudo-graviton and ghost propagators 
become [8]: 
$$\eqalignno{i \Bigl[{_{\mu \nu}} \Delta^{\rho \sigma}\Bigr](x;x') \approx 
\; &{H^2 \over 8 {\pi}^2} \; \Biggl\{ {2u'u \over {\Delta x}^2 - 
{\Delta u}^2 + 2 i \epsilon \vert {\Delta u} \vert + \epsilon^2} \;
\Bigl[{2 \delta_{\mu}^{~ (\rho} \; \delta_{\nu}^{~ \sigma)}} - 
\eta_{\mu \nu} \; \eta^{\rho \sigma} \Bigr] \cr
&- \ln \Bigl[H^2 \Bigl({\Delta x}^2 - {\Delta u}^2 + 
2 i \epsilon \vert {\Delta u} \vert + \epsilon^2\Bigr) \Bigr] \;
\Bigl[ 2{\overline \delta_{\mu}^{~(\rho}} \;
{\overline \delta_{\nu}^{~ \sigma)}} - 2 {\overline \eta_{\mu \nu}} \;
{\overline \eta^{\rho \sigma}} \Bigr] \; \Biggr\} \qquad &(13a) \cr}$$
$$\eqalignno{i \Bigl[{_{\mu}} \Delta_{\nu}\Bigr](x;x') \approx 
\; {H^2 \over 8 {\pi}^2} \; \Biggl\{ &{2u'u \over {\Delta x}^2 - 
{\Delta u}^2 + 2 i \epsilon \vert {\Delta u} \vert + \epsilon^2} \;
\eta_{\mu\nu}\cr 
&\qquad - \ln\Bigl[H^2 \Bigl({\Delta x}^2 - {\Delta u}^2 + 
2 i \epsilon \vert {\Delta u} \vert + \epsilon^2\Bigr)\Bigr] \;
{\overline \eta_{\mu \nu}} \Biggr\} &(13b) \cr}$$
Parenthesized indices are symmetrized and a bar above a Lorentz metric 
or a Kronecker delta symbol means that the zero component is projected 
out, e.g. ${\overline \eta}_{\mu \nu} \equiv \eta_{\mu \nu} + \delta_
{\mu}^{~0} \; \delta_{\nu}^{~0}$. The decoupling between functional 
dependence upon spacetime and tensor indices --- and the simplicity of 
each --- greatly facilitates calculations.

The conventional Feynman rules give ``in''-``out'' matrix elements. 
Schwinger long ago worked out a generalization which gives ``in''-``in'' 
expectation values [12]. One first implements forward evolution from the 
asymptotic past to an arbitrary time $u_0$ by the functional integral:
\footnote{*}{\tenpoint The curious integration limits are the result of
the temporal inversion which occurs when using the conformal time:
$$\int_{-\infty}^{t_0} dt \; = \;
-\int_{\infty}^{u_0} du \; (Hu)^{-1} \; = \;
\int_{u_0}^{\infty} du \; (Hu)^{-1}$$} 
$$\Bigl\langle \phi(u_0) \Bigl\vert {\rm in} \Bigr\rangle = 
\Fint [d\phi^+] \; \exp\Biggl( i\int_{u_0}^{\infty} du \; 
{\cal L}[\phi^+] \Biggr) \eqno(14a)$$
One then evolves back to the asymptotic past by means of a different 
functional integration:
$$\Bigl\langle {\rm in} \Bigl\vert \phi(u_0) \Bigl\rangle = 
\Fint [d\phi^-] \; \exp\Biggl( -i \int_{u_0}^{\infty} du \; 
{\cal L}^*[\phi^-] \Biggr) \eqno(14b)$$
The generic dummy fields $\phi^+$ and $\phi^-$ represent the pseudo-graviton
and ghosts, and are required to agree at $u_0$. To obtain expectation values 
in the presence of free vacuum at $u = H^{-1}$ we need only change the upper 
limits in the action integrals from $\infty$ to $H^{-1}$.

The associated Feynman rules are simple: vertices can either be ``$+$'' 
(conventional) or ``$-$'' (conjugated), and propagators can link any two 
kinds of fields. The three distinct sorts of propagators are:
$$\Bigl\langle 0 \Bigl\vert T\Bigl\{\psi^+_{\mu \nu}(x) \; 
\psi^+_{\rho \sigma}(x')\Bigr\} \Bigr\vert 0 \Bigr\rangle_{\rm free} = 
H^3 \sum_{\lambda, {\vec k} \neq 0} \Biggl\{ \theta({\Delta u}) \; 
\Psi_{\mu \nu} \; {\Psi'}^*_{\rho \sigma} + \theta(-{\Delta u}) \; 
\Psi^*_{\mu \nu} \; {\Psi'}_{\rho \sigma} \Biggr\} 
e^{- \epsilon \Vert {\vec k}\Vert} \eqno(15a)$$
$$\Bigl\langle 0 \Bigl\vert T\Bigl\{\psi^+_{\mu \nu}(x) \; 
\psi^-_{\rho \sigma}(x')\Bigr\} \Bigr\vert 0 \Bigr\rangle_{\rm free} = 
H^3 \sum_{\lambda, {\vec k} \neq 0} \Psi^*_{\mu \nu} \;
{\Psi'}_{\rho \sigma} \; e^{- \epsilon \Vert {\vec k} \Vert } \eqno(15b)$$
$$\Bigl\langle 0 \Bigl\vert T\Bigl\{\psi^-_{\mu \nu}(x) \; 
\psi^-_{\rho \sigma}(x')\Bigr\} \Bigr\vert 0 \Bigr\rangle_{\rm free} = 
H^3 \sum_{\lambda, {\vec k} \neq 0} \Biggl\{ \theta({\Delta u}) \;
\Psi^*_{\mu \nu} \; {\Psi'}_{\rho \sigma} + \theta(-{\Delta u}) \; 
\Psi_{\mu \nu} \; {\Psi'}^*_{\rho \sigma} \Biggr\} 
e^{- \epsilon \Vert {\vec k} \Vert} \eqno(15c)$$
(Recall that ${\Delta u} \equiv u'-u$.) Of course (15a) is just the Feynman 
propagator (13a), and (15c) is its conjugate. The ``mixed'' propagator (15b) 
can be obtained from (13a) by the replacement: $\vert {\Delta u} \vert 
\longrightarrow - {\Delta u}$. Because this mixed propagator agrees with 
(15a) for $\Delta u < 0$ and with (15c) for $\Delta u > 0$, and because the 
imaginary part of any propagator is zero for spacelike separation, it is 
simple to show that there will be destructive interference between the 
various ``$+$'' and ``$-$'' interaction vertices contributing to a single 
observation unless all vertices lie within the past lightcone of the 
observation point.

\vskip 1cm
\centerline {\bf 3. The result}

We actually computed the amputated expectation value of $\kappa \psi_
{\mu \nu} (u,{\vec x})$, then attached the external leg and multiplied 
by $\Omega^2$ to obtain the corrected background. Because the initial 
condition, the background, and the dynamics are rotationally and 
translationally invariant, the amputated 1-point function can be 
expressed in terms of just two functions of $u$ [4]:
\footnote{*}{\tenpoint The gauged-fixed kinetic operator 
$D_{\mu \nu}^{~~\rho \sigma}$ is most conveniently expressed in terms
of the kinetic operator ${\rm D}_A \equiv \Omega (\partial^2 + 
\frac2{u^2}) \Omega$ for a massless, minimally coupled scalar and 
the kinetic operator ${\rm D}_B = {\rm D}_C \equiv \Omega \> 
\partial^2 \Omega$ for a conformally coupled scalar [9]: 
$$D_{\mu \nu}^{~~ \rho \sigma} \equiv \Bigl[
\frac12 {\overline \delta}_{\mu}^{~(\rho} \; 
{\overline \delta}_{\nu}^{~\sigma)} - 
\frac14 \eta_{\mu \nu} \; \eta^{\rho\sigma} - 
\frac12 \delta_{\mu}^{~0} \; \delta_{\nu}^{~0} \; 
\delta_0^{~\rho} \; \delta_0^{~\sigma} \Bigr] {\rm D}_A + 
\delta_{(\mu}^{~~0} \; {\overline \delta}_ {\nu)}^{~~(\rho} \; 
\delta_0^{~\sigma)} \; {\rm D}_B + 
\delta_{\mu}^{~0} \; \delta_{\nu}^{~0} \; 
\delta_0^{~\rho} \; \delta_0^{~\sigma} \; {\rm D}_C$$}
$$D_{\mu \nu}^{~~\rho \sigma} \; \Bigl\langle 0 \Bigl\vert \;
\kappa \psi_{\rho \sigma}(x) \; \Bigr\vert 0 \Bigr\rangle =
a(u) \; {\overline \eta}_{\mu \nu} + c(u) \;
\delta^0_{~\mu} \delta^0_{~\nu} \eqno(16)$$
Neither $a(u)$ nor $c(u)$ is independent of the choice of gauge --- 
only the combination of them which goes to make up $H_{\rm eff}(t)$ 
is. A necessary condition for $H_{\rm eff}(t)$ to receive a non-zero 
contribution at late times ($u \longrightarrow 0^+$) is that either $a(u)$ 
or $c(u)$ must grow at least as fast as $u^{-4}$ [4]. In fact the only 
interest is in faster growth, since $u^{-4}$ behavior could be absorbed 
into a renormalization of $\Lambda$. The fastest growth possible at any 
order in perturbation theory is powers of $\ln(H u)$ over $u^4$ [4,13]. 
These logarithms can come either from integrating the factors of $\Omega = 
(Hu)^{-1}$ one can see on the interaction vertices (7), or from the 
logarithm term of an undifferentiated  propagator (13). The physical
origin of the former effect is just the growth of the invariant volume 
of the past lightcone; the latter effect derives from the increasing 
correlation of the vacuum at constant invariant separation as inflation 
proceeds.

\vskip -1.5cm

\centerline{\psfig{figure=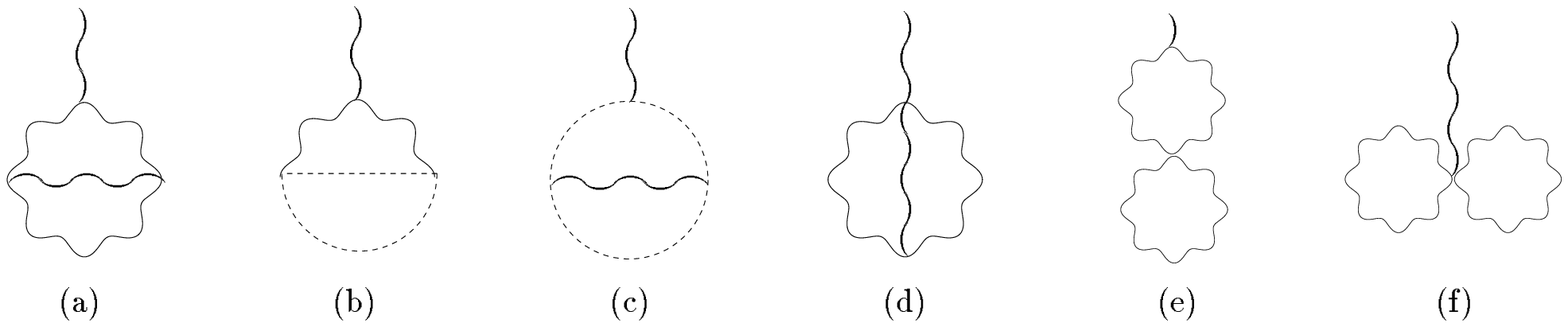,height=21cm,bbllx=0bp
,bblly=0bp,bburx=596bp,bbury=843bp,rheight=5.5cm,rwidth=12.7cm}}

{\bf Fig.~1:} {\ninepoint Two-loop contributions to the background 
geometry. Gravitons reside on wavy lines and ghosts 
\vskip -8pt \noindent \hglue 2.40truecm on segmented lines.}

\vskip 0.5cm

There are no logarithms at one loop because amputated one-loop tadpoles 
involve no integrations. In fact, it turns out that the $u^{-4}$ threshold 
is not even reached. The six diagrams of Fig.~1 contribute to the two-loop 
tadpole. Diagram (f) can give no logarithms because it involves no 
integrations, and diagram (e) is entirely canceled by the counterterm 
needed to renormalize its coincident inner loop. Diagram (d) contributes 
two logarithms because the volume factor from the integration can be 
enhanced by the single possible undifferentiated logarithm which can 
survive from the three propagators. A single undifferentiated logarithm 
can also survive from the four propagators in diagrams (a-c), and one 
might think that the two volume factor integrations could be enhanced to 
produce terms of order $u^{-4} \; \ln^3(H u)$. However, it turns out that 
whenever an undifferentiated logarithm survives, the structure of the 
associated factors always precludes triple logarithms [14].
\footnote{*}{\tenpoint Although the reason for this is now clear, it was
not obvious from the way in which the calculation had to be broken up in order
to be performed efficiently. Indeed, the result did seem to contain triple
logarithm terms [15] due to a factor of two error in performing the angular
integration for a rare special class of denominators. Now that the result has
been corrected and thoroughly checked one can regard the complete cancellation
of the triple logarithm terms as a powerful additional check on accuracy.}
 
We have reported on the details of the calculation elsewhere [14]. The 
result is:
$$a(u) = H^{-2} \Bigl({\kappa H \over 4 \pi u}\Bigr)^4 \Biggl\{ 
\Bigl(- \frac{1795}{9} + \frac{604}{9} + \frac{320}{3} - \frac{52}{3} 
\Bigr) \ln^2(Hu) + {\rm (subleading)}\Biggr\} + O(\kappa^6) \eqno(17a)$$
$$c(u) = H^{-2} \Bigl({\kappa H \over 4 \pi u}\Bigr)^4 \Bigl\{ 
\Bigl(\frac{1157}{3} - \frac{800}{3} - {\scriptstyle 112} + 
{\scriptstyle 8} \Bigr) \ln^2(Hu) + {\rm (subleading)}\Bigr\} + 
O(\kappa^6) \eqno(17b)$$
The four numerical coefficients in $a(u)$ and $c(u)$ represent
the individual contributions of the diagrams (a) through (d) of Fig.~1 
respectively. The external leg is a retarded Green's function in
Schwinger's formalism. The procedure for attaching this is straightforward 
[4], and the invariant interval: 
$$\Bigl\langle 0 \Bigl\vert \; g_{\mu \nu}(t,{\vec x}) \;
dx^{\mu} dx^{\nu} \; \Bigr\vert 0 \Bigr\rangle = \Omega^2 \Bigl\{ 
- \Bigl[1 - C(u)\Bigr] \; du^2 + \Bigl[1 + A(u)\Bigr] \; 
d{\vec x} \cdot d{\vec x}\Bigr\} \eqno(18a)$$
acquires the following leading result for small $u$:
$$A(u) = \Bigl({\kappa H \over 4 \pi}\Bigr)^4 \Bigl\{ 
\Bigl(\frac{7180}{81} - \frac{2416}{81} - \frac{1280}{27} + \frac{208}{27} 
\Bigr) \ln^3(Hu) + {\rm (subleading)}\Bigr\} + O(\kappa^6) 
\eqno(18b)$$
$$C(u) = \Bigl({\kappa H \over 4 \pi}\Bigr)^4 \Bigl\{ 
\Bigl(\frac{319}{6} + \frac{49}{3} - {\scriptstyle 52} + 
{\scriptstyle 11} \Bigr) \ln^2(Hu) + {\rm (subleading)}\Bigr\} + 
O(\kappa^6) \eqno(18c)$$
Comparing this with expressions (2) and (4) gives the following formula 
for the effective Hubble constant:
$$\eqalignno{H_{\rm eff}(t) &= {H \over \sqrt{1 - C(u)}} \; \Bigr\{
1 - \frac12 u {d \over du} \ln\Bigl[1 + A(u)\Bigr]\Bigr\} &(19a) \cr
&= H \Bigg\{1 - \Bigl({\kappa H \over 4 \pi}\Bigr)^4 \Bigl[ \Bigl(
\frac{4309}{54} - \frac{1649}{27} - \frac{172}{9} + \frac59 \Bigr) 
(H t)^2 + {\rm (sub.)}\Bigr] + O(\kappa^6)\Biggr\} &(19b) \cr
&= H \Bigg\{1 - \Bigl({\kappa H \over 4 \pi}\Bigr)^4 \Bigl[ \; 
\frac16 \; (H t)^2 + {\rm (subleading)}\Bigr] + 
O(\kappa^6)\Biggr\} &(19c) \cr}$$
Note that each of the four diagrams contributes with the sign we might 
have expected. The pure graviton diagrams (a) and (d) slow inflation 
because the negative gravitational interaction energy reduces the vacuum 
energy. On the contrary,  diagrams (b) and (c) have the opposite sign 
because ghost loops remove unphysical graviton modes from (a) and (d).

It is worth noting that only one other two-loop result has been obtained in 
quantum gravity, and this was limited to the ultraviolet divergent part for 
zero cosmological constant [16]. There is little doubt that our formalism is
consistent and that the basic reduction procedure is correct, however, it is 
legitimate to worry about the accuracy of implementation in a calculation of
this complexity. We have subjected our work to every available check [14,17] 
and we are reasonably sure it is correct. We nonetheless feel that computer
calculations of this scale should be regarded as experiments which can and 
should be independently verified before being completely trusted.\footnote{*}{
\tenpoint We will provide copies of the programs used and the intermediate data
generated to anyone who wishes to examine them.}

\vskip 1cm
\centerline {\bf 4. Interpretation}

A simple consequence of (19) is that quantum gravity induces an effective 
stress-tensor which obeys the equation of state for negative vacuum energy, 
at least in the initial stages of relaxation. To see this we evaluate the 
classical field equations for the full quantum solution (2) and define the 
stress tensor as $(8 \pi G)^{-1}$ times the deficit:
$$G_{\mu \nu} + \Lambda g_{\mu \nu} \equiv 8 \pi G \; T_{\mu \nu} 
\eqno(20)$$ 
The induced energy and pressure can be expressed in terms of $H_{\rm eff}(t)$
as follows:
$$\rho(t) = \frac{6}{\kappa^2} \Bigl[ H^2_{\rm eff}(t) - H^2\Bigr] 
\eqno(21a)$$ 
$$p(t) = - \rho(t) - \frac{4}{\kappa^2} {\dot H}_{\rm eff}(t) \eqno(21b)$$
For (19c) we get:
$$\rho(t) = - G H^6 \; \Bigl\{ \Bigl[ \frac{1}{8 \pi^3} (H t)^2 + 
{\rm (subdominant)}\Bigr] + O(\kappa^2 H^2)\Bigr\} \eqno(22)$$ 
where $p = - \rho + {\rm (subdominant)}$. Note that whenever $Ht$ is large 
enough to compensate for the small prefactor, the subdominant terms are 
truly insignificant. 

The most important consequence of our result is that quantum gravitational 
processes slow the rate of inflation by an amount which becomes 
non-perturbatively large at late times. Of course this had to be so in 
view of the previously cited ``in''-``out'' results [3,4,8], but it is 
nice to have a proof, and even nicer to know the rate. If there were no 
further corrections we could estimate the number of e-foldings needed to 
extinguish inflation as:
$$Ht \sim (\kappa H)^{-2} \eqno(23)$$
For GUT scale inflation this would be about $10^{12}$; the corresponding 
result for electroweak inflation would be about $10^{68}$. These estimates 
are not altered by higher loop corrections because the leading contribution 
at $\ell$ loops has the form:
$$-\# (\kappa H)^{2 \ell} \Bigl(\ln(Hu)\Bigr)^{\ell} \eqno(24)$$
Of course the number of coupling constants is simple to compute. To get
the temporal dependence note that dimensional analysis requires each factor 
of $\ln(u)$ to be associated either with $\ln(H)$ or $-\ln(\epsilon)$. The
factors of $\ln(\epsilon)$ come from ultraviolet divergences, and we know
that there can be at most $\ell$ of them in an $\ell$-loop graph. In 
principle one could get factors of $\ln(H)$ from the upper limits of the
conformal time integrations, but dimensional analysis reveals that the
integrands fall off too rapidly at large conformal times for this. 
Therefore, any factor of $\ln(H)$ must come from the logarithm term of 
an undifferentiated propagator such as (13a) or (13b). But this means that 
the relevant part of the associated line is just a constant --- $\ln(H)$ 
--- {\it so one of the loops is cut and the maximum number of ultraviolet 
logarithms is one fewer.} There can be at most $\ell-1$ undifferentiated 
logarithms in an $\ell$-loop tadpole, so such a graph can contribute $k$ 
factors of $\ln(H)$ and up to $\ell-k$ factors of $\ln(\epsilon)$ for 
$k=0,1,\dots,\ell-1$. Hence there can be at most $\ell$ factors of $\ln(u)$,
and the strongest possible growth at late times is indeed (24).

We can infer the sign of the higher loops effects by using Feynman's tree 
theorem [18] and appealing to common sense about classical gravitation. 
The tree theorem is a procedure for decomposing loops into sums of on-shell 
tree diagrams. The decomposition comes from rearranging the mode expansions 
of the various propagators (15a-c). For example, the ``$++$'' propagator 
can be written as:
$$\eqalignno{&\Bigl\langle 0 \Bigl\vert T\Bigl\{\psi^+_{\mu \nu}(x) \; 
\psi^+_{\rho \sigma}(x')\Bigr\} \Bigr\vert 0 \Bigr\rangle_{\rm free} = 
H^3 \sum_{\lambda, {\vec k} \neq 0} \Biggl\{ \theta({\Delta u}) \; 
\Psi_{\mu \nu} \; {\Psi'}^*_{\rho \sigma} + 
\theta(-{\Delta u}) \; \Psi^*_{\mu \nu} \; {\Psi'}_{\rho \sigma} \Biggr\} 
e^{- \epsilon \Vert {\vec k}\Vert} \cr
&= \theta({\Delta u}) \; H^3 \sum_{\lambda, {\vec k} \neq 0} 
\Biggl\{ \Psi_{\mu \nu} \; {\Psi'}^*_{\rho \sigma} - 
\Psi^*_{\mu \nu} \; {\Psi'}_{\rho \sigma} \Biggr\} 
e^{- \epsilon \Vert {\vec k}\Vert} + 
H^3 \sum_{\lambda, {\vec k} \neq 0} \Psi^*_{\mu \nu} \; 
{\Psi'}_{\rho \sigma} \; e^{- \epsilon \Vert {\vec k}\Vert} &(25) \cr}$$
The first sum is just the retarded propagator, while the second sum is the
operator for on-shell propagation from $x'$ to $x$. The mixed propagator 
(15b) is already on-shell, and the ``$--$'' propagator (15c) is just the 
conjugate of (25). Analogous results apply as well for the various ghost 
propagators. The tree decomposition of a loop is obtained by expanding the 
various propagators as retarded plus on-shell ones. The term with all 
retarded propagators vanishes --- except for local terms from the singular 
coincidence limit, which is in the hard ultraviolet and of no concern to 
us. The remaining terms are sums of on-shell trees. 

The tree decomposition allows us to regard deviations from the classical 
background as the response to an ensemble of on-shell gravitons. (Ghosts 
cancel against unphysical graviton polarizations as a consequence of 
decoupling.) This interpretation is advantageous because although isolated 
classical gravitons carry positive energy, the interaction energy between 
them is negative, at least for long wavelengths. In fact we have already 
encountered both terms at lowest order: the positive graviton zero-point 
energy comprises the one-loop tadpole, while the negative gravitational 
interaction energy gives the dominant infrared behavior of the two-loop 
tadpole. Like the one-loop tadpole, the energy of isolated gravitons is 
ultra-local, ultraviolet divergent and completely subsumed into local 
counterterms. The gravitational interaction energy is not local because 
it represents the interaction between diffuse sources. Its pressure obeys 
the equation of state for vacuum energy, $p = - \rho$, because causality 
limits the range of interaction to a Hubble radius. The energy density 
must therefore be independent of the much larger, total volume. Hence the 
total energy is $U = \rho V$ and we have:
$$p = - {\partial U \over \partial V} = - \rho \eqno(26)$$
We can even understand the prefactor of $G H^6$ in (22) by considering the 
Newtonian gravitational interaction energy due to the zero-point energy 
of a mode of wavelength $H^{-1}$ confined to a Hubble volume $H^{-3}$. 

The time dependence of the effect --- which is why it cannot be absorbed 
into the initial cosmological constant --- can be roughly understood by 
considering the response of the background zero mode to a single graviton 
of initial momentum ${\vec k} \neq 0$. There is little response initially 
because the spatial variation of the graviton gives it a weak overlap with 
the zero mode. Inflation redshifts the physical momentum to $H u \; 
{\vec k}$. As the graviton flattens out the zero mode feels it more 
strongly, however, the process is cut off by causality when the graviton's 
wavelength redshifts beyond the causal horizon of $H^{-1}$. We expect 
the maximum effect when:
$$Hu \; \Vert {\vec k} \Vert \sim H \qquad \Longrightarrow 
\qquad u \sim \Vert {\vec k} \Vert^{-1} \eqno(27)$$
If the radiation ensemble were cut off at high momentum then we would 
see the ensemble's gravitational interaction energy slow inflation until 
the shortest wavelength had been redshifted beyond the causal horizon, 
after which there would be little effect. A locally de Sitter background 
is stable on the classical level [10] because any classically well defined 
metric must possess such a cutoff. The fact that quantum gravity continues 
to show an effect at late times derives from the absence of a cutoff in 
the tree decomposition, which absence is itself a consequence of the 
uncertainty principle. We see an {\it ever increasing} effect because 
the number of modes just redshifting past the horizon grows like $\Vert 
{\vec k} \Vert^2 \sim u^{-2}$.

\vfill\eject
\centerline {\bf 5. Conclusions}

To sum up, we have proposed that inflation happens for no other reason 
than that the bare cosmological constant is large and positive --- there
is no need for scalars. The observed universe derives from a patch of 
about one Hubble volume which began inflation when its local temperature 
dropped below the scale $M = (H/\kappa)^{1/2}$. The reason our universe 
is not inflating today --- at least not rapidly --- is that infrared 
processes in quantum gravity tend to screen the bare cosmological 
constant. The irresistible force of the effect comes from the fact 
that it scales as powers of the invariant volume of the past lightcone, 
which grows without bound as evolution continues. The slowness of the 
effect derives from the fact that it is a gravitational process, and 
only the causal and coherent superposition of interactions over an 
enormous invariant volume can overcome the weakness of the natural 
coupling constant, $\kappa H \ltwid 10^{-6}$.

All experimentally confirmed matter quanta are either massive or else 
conformally invariant at the classical level, so they give only 
negligibly small corrections to the classical geometry of an inflating 
universe. Although certain conjectured light quanta may be competitive 
with the graviton for a time, in the end only gravity is left. This 
means that quantum gravity makes {\it unique} predictions for the 
extinction of inflation. Unfortunately, the most interesting of these 
are not easy to obtain because they occur after perturbation theory 
has broken down. In particular, the last sixty e-foldings govern the 
magnitude and spectrum of observable density perturbations, and the 
very end of inflation should tell us what reheating temperature was 
reached.

Without some control of the regime in which the infrared effect 
becomes large, it is hard to make even qualitative arguments. The 
onset of the strong effect may be sudden or slow; and the approach 
of $H_{\rm eff}(t)$ to zero may be monotonic or involve an ``overshoot" 
followed by damped oscillations between $\Lambda > 0$ and $\Lambda 
< 0$ phases. For instance, if the onset is sudden with subsequent 
oscillations the reheating temperature should be comparable with 
the initial scale of inflation, ${\rm T_{RH}} \sim (H/\kappa)^{1/2}$.
But if the approach is slow and monotonic we expect a lower value 
of ${\rm T_{RH}}$.

At least in the perturbative regime, the adiabatic analysis can be
shown to underestimate the strength of the effect. For if we divide
the time interval into $N$ steps $dt = (t/N)$ and evolve forward by 
reseting at the beginning of every step the bare cosmological constant 
to equal its effective value at the end of the previous step, after 
$N$ iterations we obtain:
$$H_{\rm eff}^{(N)}(dt) = 
H \Bigg\{1 - \Bigl({\kappa H \over 4 \pi}\Bigr)^4 \Bigl[ \; 
\frac16 \; \frac{(H t)^2}{N} + {\rm (subleading)}\Bigr] + 
O(\kappa^6)\Biggr\} \eqno(28)$$
This result does not agree with the $H_{\rm eff}(t)$ given by (19c)
and gives a weaker effect. 

What happens beyond perturbation theory is an open problem. Screening 
depends upon the strength of the interaction in the past lightcone,
the strength of correlations from the past with the present and the
accessible volume of the past lightcone. Notice that the first two 
decline as inflation slows while the third increases. The delicate 
balance between the different ways the above three factors evolve 
determines what will actually happen. It is quite possible that the
screening is never complete. If so, part of our current expansion is 
residual inflation, and the time dependence of this component may play 
a crucial role in resolving, among other things, the apparent problem 
with the age of the universe. Therefore, there may even be observable 
effects in the present epoch.

\vskip 1cm
\centerline{ACKNOWLEDGEMENTS}

We wish to thank T. J. M. Zouros for the use of his SUN SPARC station, 
and S. Deser for his encouragement and support during this project. 
We also acknowledge discussions with I. Antoniadis, C. Bachas and J.
Iliopoulos. One of us (RPW) thanks the University of Crete and the 
Theory Group of FO.R.T.H. for their hospitality during the execution 
of this project. This work was partially supported by DOE contract 
86-ER40272, by NSF grant 94092715 and by EEC grant 933582.

\vskip 1cm
\centerline{REFERENCES}

\item{[1]} A. H. Guth, {\sl Phys. Rev.} {\bf D23} (1981) 347.
\hfill\break  
E. W. Kolb and M. S. Turner, {\it The Early Universe} 
(Addison-Wesley, Redwood City, CA, 1990).
\hfill\break
A. D. Linde, {\it Particle Physics and Inflationary Cosmology} (Harwood,
Chur, Switzerland, 1990).

\item{[2]} A. Sandage, {\sl Observatory} {\bf 88} (1968) 91.

\item{[3]} N. C. Tsamis and R. P. Woodard, {\sl Phys. Lett.} {\bf B301} (1993) 
351.

\item{[4]} N. C. Tsamis and R. P. Woodard, {\sl Ann. Phys.} {\bf 238} (1995) 1.

\item{[5]} N. P. Myrhvold, {\sl Phys. Rev.} {\bf D28} (1983) 2439.
\hfill\break
E. Mottola, {\sl Phys. Rev.} {\bf D31} (1985) 754.

\item{[6]} I. Antoniadis and E. Mottola, {\sl Phys. Rev.} {\bf D45} (1992) 2013.
\hfill\break 
E. Elizalde and S. D. Odintsov, {\sl Phys. Lett.} {\bf B334} (1994) 33.

\item{[7]} L. H. Ford, {\sl Phys. Rev.} {\bf D31} (1985) 710.

\item{[8]} N. C. Tsamis and R. P. Woodard, {\sl Class. Quantum Grav.} {\bf 11} 
(1994) 2969.

\item{[9]} N. C. Tsamis and R. P. Woodard, {\sl Commun. Math. Phys.} {\bf 162} 
(1994) 217.

\item{[10]} S. Deser and L. F. Abbott, {\sl Nucl. Phys.} {\bf B195} (1982) 76.
\hfill\break
P. Ginsparg and M. J. Perry, {\sl Nucl. Phys.} {\bf B222} (1983) 245.

\item{[11]} N. C. Tsamis and R. P. Woodard, {\sl Phys. Lett} {\bf B292} (1992) 
269.

\item{[12]} J. Schwinger, {\sl J. Math. Phys.} {\bf 2} (1961) 407; {\it 
Particles, Sources and Fields} (Addison-Wesley, Reading, MA, 1970).

\item{[13]} A. D. Dolgov, M. B. Einhorn and V. I. Zakharov, {\sl Phys. Rev.}
{\bf D52} (1995) 717.

\item{[14]} N. C. Tsamis and R. P. Woodard, ``The Quantum Gravitational 
Back-Reaction On Inflation,'' {\it hep-ph/9602316}, to appear in {\sl
Annals of Physics}.

\item{[15]} R. P. Woodard, in {\it Quantum Infrared Physics, Proceedings 
of the Workshop at the American University of Paris, 6--10 June 1994}, 
ed.  H. M. Fried and B. M\"uller (World Scientific, Singapore, 1995) 
pp. 450-459.
 
\item{[16]} M. Goroff and A. Sagnotti, {\sl Phys. Lett.} {\bf B160} (1986) 81;
{\sl Nucl. Phys.} {\bf B266} (1986) 709.

\item{[17]} N. C. Tsamis and R. P. Woodard, {\sl Phys. Rev.} {\bf D15} (1996) 
2621.

\item{[18]} R. P. Feynman, {\sl Acta Phys. Pol.} {\bf 24} (1963) 697; in 
{\it Magic Without Magic}, ed. J. Klauder (Freeman, New York, 1972) 355.

\bye